 \documentclass[10pt,preprint]{aastex}  
 \usepackage{natbib}
\def\gapprox{\lower.4ex\hbox{$\;\buildrel >\over{\scriptstyle\sim}\;$}}
\def\lapprox{\lower.4ex\hbox{$\;\buildrel <\over{\scriptstyle\sim}\;$}}

\shortauthors{ASCHWANDEN}
\shorttitle{Braiding and Nanoflaring}

\begin{document}

\title{		The Minimum Energy Principle Applied to Parker's Coronal 
		Braiding and Nanoflaring Scenario}

\author{        Markus J. Aschwanden$^1$}

\affil{         $^1)$ Lockheed Martin,
                Solar and Astrophysics Laboratory,
                Org. A021S, Bldg.~252, 3251 Hanover St.,
                Palo Alto, CA 94304, USA;
                e-mail: aschwanden@lmsal.com }

\and

\author{	A. A. van Ballegooijen$^2$}

\affil{		$^2)$ 5001 Riverwood Avenue, Sarasota, FL 34231, USA}

\begin{abstract}
Parker's coronal braiding and nanoflaring scenario predicts
the development of tangential discontinuities and highly misaligned
magnetic field lines, as a consequence of random buffeting of their
footpoints due to the action of sub-photospheric convection. The
increased stressing of magnetic field lines is thought to become 
unstable above some critical misalignment angle and to result into 
local magnetic reconnection events, which is generally referred to as 
Parker's ``nanoflaring scenario''. In this study we show that the 
{\sl minimum (magnetic) energy principle} leads to a bifurcation of 
force-free field solutions for helical twist angles at 
$|\varphi(t)| = \pi$, which prevents the build-up of arbitrary large free 
energies and misalignment angles. The minimum energy principle predicts 
that neighbored
magnetic field lines are almost parallel (with misalignment angles of
$\Delta \mu \approx 1.6^\circ-1.8^\circ$), and do not reach
a critical misalignment angle prone to nanoflaring. Consequently,
no nanoflares are expected in the divergence-free and force-free
parts of the solar corona, while they are more likely to occur
in the chromosphere and transition region.
\end{abstract}


\section{	INTRODUCTION					}

The sub-photospheric layers of the Sun exhibit vigorous convective motions, 
which are driven by a negative vertical temperature gradient $dT/dh < 0$ 
due to the B\'{e}nard-Rayleigh instability (Lorenz 1963). The convective 
motion of the sub-photospheric fluid is a self-organizing process that 
creates granules with a characteristic size $\ell \approx 1000$ km and 
horizontal velocity $v \approx 2$ $\rm km ~ s^{-1}$. It follows that the 
convective motions have a typical time scale $\tau = \ell / v \approx 500$ s, 
or 7 minutes, which matches the observed life time of granules. The question 
now arises how the solar photosphere affects the coronal magnetic field. 
In Parker's theory for solar coronal heating (Parker 1972, 1983), a coronal 
loop is perturbed by small-scale, random motions of the photospheric 
footpoints of the coronal field lines, driven by sub-photospheric convective 
flows. According to Parker, these motions lead to twisting and braiding of 
the coronal field lines. As the magnetic field evolves, thin current sheets 
are formed in the corona, and energy is dissipated by magnetic reconnection. 
The reconnection is likely to occur in a burst-like manner as a series of
``nanoflares'' (Parker 1988). Parker's magnetic braiding scenario has 
become the main paradigm for coronal heating in active regions, and the 
nanoflare heating theory has been widely adopted (see reviews by 
Cargill 2015; Klimchuk 2015; Aschwanden 2005). 
However, the response of the coronal magnetic 
field to random footpoint motions is a complex physical problem, and the 
implications of Parker's theory are not yet fully understood.

Parker (1983, 1988) created several cartoons that depict how the random 
motion of coronal magnetic field lines (or lines of force) wind around 
neighbored, less-twisted (or untwisted) field lines (Fig.~1), either for 
dipolar field lines (Fig.~1a) or in the form of stretched-out flux tubes 
(Fig.~1b). The winding motion has been aptly called ``coronal loop braiding," 
although it has never been demonstrated that a braided field can be formed 
by random motions of the photospheric footpoints. Some cartoons display a 
single braided field line among a set of untwisted (straight) field lines 
(Fig.~1c), or a single flux tube surrounded by a set of untwisted (straight) 
flux tubes (Fig.~1d). The magnetic braiding model predicts that the 
misalignment angle between a braided field line and a neighboring 
less-braided field line will increase with time, building up magnetic 
stress until a reconnection event is triggered, which causes some or all 
of the magnetic energy to be released.

When a coronal loop system is viewed from the side, it generally shows a 
collection of nearly parallel threads that do not cross each other, and 
neighboring threads are co-aligned to within a few degrees. The threads 
are believed to follow the magnetic field ${\bf B} ({\bf r})$, so the 
direction of ${\bf B}$ in neighboring threads must be co-aligned to 
within a few degrees. In many cases the average direction of the threads 
is consistent with that predicted by potential field models. Therefore, 
the observations provide little evidence for the presence of braided 
fields with misalignment angles of about 20 degrees relative to the 
potential field, as predicted by Parker (1983). The observations are 
consistent with the much simpler geometries rendered in Figure~2, 
which shows a dipolar flux tube system with slightly twisted (Fig.~2a,b) 
or a strongly twisted (Fig.~2c,d) flux tubes.

In this paper we consider the assumptions of Parker's braiding model and 
its suitability to explain the observed coronal loops. We argue that the 
model is not internally consistent because it predicts that thin current 
sheets will form quickly, but reconnection is assumed to be postponed until 
the braided field is fully formed. It is not clear that a braided field 
will form because the timescale for the formation of thin current sheets 
(a few minutes) is much shorter than the time needed to build up the 
braided field (hours). We propose an alternative ``solution'' to the Parker 
problem in which reconnection causes the magnetic field to remain close 
to a minimum energy state in which the field lines do not deviate strongly 
from straight lines (also see Rappazzo \& Parker 2013; Rappazzo 2015). 
According to this scenario, the relative displacements of the footpoint 
at the two boundary plates in the Parker model are not much larger than 
the correlation length $\ell$ of the footpoint motions. We also compare 
the predictions of the minimum energy scenario with results from MHD 
simulations, and discuss the role of magnetic braiding in coronal heating. 
We find that quasi-static braiding driven by granule-scale footpoint 
motions cannot provide enough energy to heat active-region loops to 
the observed temperatures and densities.

\section{RECONNECTION IN THE MAGNETIC BRAIDING MODEL}

Parker (1972, 1983) considered a simplified model for a coronal loop in 
which an initially uniform field is contained between two parallel plates, 
which represent the photosphere at the two ends of the loop. The coronal 
plasma is assumed to be highly conducting, so the magnetic field is nearly 
``frozen'' into the plasma. The loop is anchored in the photosphere at its 
two ends, and its length $L$ is assumed to be much larger than its width, 
so the curvature of the loop can be neglected. A cartesian coordinate 
system $(x,y,z)$ is used with the $z$-axis along the (straigthened) loop. 
The planes $z=0$ and $z=L$ represent the photosphere at the two ends of 
the loop, and the ``corona'' is the region in between these planes 
($0 < z < L$). At time $t = 0$ the magnetic field is assumed to be a 
uniform potential field, ${\bf B} ({\bf r},0) = B_0 \hat{\bf z}$, 
where $B_0$ is the field strength. The field is perturbed by random 
footpoint motions at $z = 0$ and $z = L$, which cause random twisting 
and braiding of the coronal field lines. In this model all details of 
the lower atmosphere are neglected, and footpoint motions with 
velocities of 1 -- 2 $\rm km ~ s^{-1}$ are applied directly at the 
coronal base. This simplified version is very useful because it is well 
defined mathematically and can be studied in great detail.

In the Parker problem the footpoint velocities are assumed to be 
incompressible ($\nabla_\perp \cdot {\bf v}_\perp = 0$), and vary 
randomly in space and time. The auto-correlation time $\tau$ is 
assumed to be large compared to the Alfv\'{e}n travel time $L/v_A$ 
in the corona. Then the coronal field will remain close to a 
force-free state in which the Lorentz force nearly vanishes, 
${\bf j} \times {\bf B} \approx 0$, where ${\bf j} ({\bf r},t)$ 
is the electric current density (quasi-static braiding). The 
magnetic stresses will build up over time, and in some locations 
the stress will become so large that some energy is released by 
small-scale reconnection. Hence, we expect that the magnetic field 
will evolve quasi-statically most of the time, but this evolution 
will sometimes be interrupted by reconnection events where magnetic 
free energy is converted into heat of the plasma.

Parker (1983) argued on the basis of the observed coronal heating rates 
that, starting from an initial uniform field, the photospheric footpoints 
motions must continue for a period of about 8 hours before reconnection 
can start removing some of the energy. This long energy buildup time is 
needed because the transverse component of the coronal field must build 
up to a significant fraction of the parallel component, otherwise the 
rate of energy input into the corona (the Poynting flux at the coronal 
base) is too low compared to the plasma heating rate. During this buildup 
phase strong electric currents develop in between the mis-aligned coronal 
flux tubes.

In the following we argue that the concept of magnetic braiding, although 
convincing in cartoon representations (Fig.~1), may not be the correct 
solution to the Parker problem. On the one hand, the braiding model predicts 
the ``spontaneous formation of tangential discontinuities'' in the magnetic 
field, i.e., the formation of thin current sheets (Parker 1972). Although 
the validity of Parker's (1972) argument has been questioned (e.g., 
van Ballegooijen 1985), it is expected that thin current sheets will 
develop in the Parker problem on the timescale of the solar granulation 
(at most a few minutes). This is certainly the case when discrete flux 
tubes are shuffled about as indicated in Fig.~1d. On the other hand, the 
model assumes that magnetic reconnection somehow does not occur at these 
current sheets until the braided field is fully formed. This is unlikely 
because the corona has a small but finite resistivity, and thin current 
sheets are expected to be unstable to resistive instabilities such as 
tearing modes (e.g., Tenerani et al. 2016). Therefore, thin current 
sheets cannot exist for the time needed to build up the braided field, 
which requires about 8 hours (Parker 1983). Hence, the magnetic braiding 
model is not internally consistent: it assumes magnetic reconnection 
is somehow prevented until the braided field has fully formed, then 
reconnection is ``turned on'' to explain the release of energy in 
nanoflares. Instead, one would expect that reconnection occurs whenever 
thin current sheets are present. Magnetic energy will then be released 
well before the braided field is fully formed.

Other analytical studies predicted a somewhat different scenario for what 
happens in the Parker problem. Van Ballegooijen (1985) proposed that the 
random footpoint motions cause a ``cascade" of magnetic energy towards 
smaller spatial scales, even though the field evolves through a series 
of force-free states and the plasma is not turbulent in the traditional 
sense. This cascade model predicts the formation of thin current layers 
on a time scale of a few times the dynamical time of the footpoint motions 
(a few minutes). This is larger than the time for the ``spontaneous 
formation of tangential discontinuities" (Parker 1972), which presumably 
occurs on the Alfv\'{e}n time scale ($L / v_A \sim 30$~s), but is much 
shorter than the 8-hour period needed to form a braided field of sufficient 
complexity (Parker 1983). Strongly braided fields do not develop in the 
cascade model, and the predicted dissipation rates are a factor 10 to 40 
lower than the observational requirements (van Ballegooijen 1986). 
Therefore, the cascade model fails as a theory for coronal heating, but 
it may give a more accurate description of what happens in the Parker 
problem when reconnection is allowed to occur.

\section{THE MINIMUM ENERGY SCENARIO}

The above considerations lead us to propose a somewhat different solution 
of the Parker problem in which magnetic reconnection is assumed to occur 
so frequently that the magnetic field lines never deviate very far from 
straight lines connecting the photospheric footpoints. The concept is 
first illustrated using a helically twisted cylindrical flux tube, then 
is generalized for random footpoint motions.

\subsection{Helically Twisted Magnetic Fields}

We use a parameterization of the 3-D magnetic field in terms of
helically twisted cylindrical flux tubes, for which analytical 
divergence-free and force-free solutions of the magnetic field are
known, either for a straight cylinder (Priest 1982), or in form of
the {\sl vertical-current approximation nonlinear force-free field 
(VCA-NLFFF)} model (Appendix A), which accommodates the full 3-D 
geometry of the spherical solar surface and the curvature of loops, 
being accurate to second order of $\alpha$ (Aschwanden 2013a).

We consider two magnetic field lines, one being a braiding
field line (solid linestyle in Fig.~3 top), and the other one a 
less-braided (straight) field line (dashed linestyle in Fig.~3 top).
The random footpoint motion at $L=0$ is represented, without loss
of generality, by a circular motion at the photospheric level (z=0).
The footpoint rotation moves the footpoint position from its initial
location at $(x_1, y_1, z_1=0)$ (Fig.~3a) to rotation angles of
$\varphi=\pi/2$ (Fig.~3b), 
$\varphi=\pi$ (Fig.~3c), 
$\varphi=\pi (3/2)$ (Fig.~3d), 
and finally back to the identical position where we started
$(x_1, y_1, z_1=0)$ (Fig.~3e). The geometry of the twisted field line 
(solid curve in Fig.~3 top) results into a helix rotated by one full turn, which
corresponds to a divergence-free and force-free solution in
cylindrical or spherical coordinates (Appendix A, Eqs.~A3-A6). 
We can define the angle between the twisted
field line (at radius $r$) and an untwisted field line (at the
cylindrical symmetry axis) with a misalignment angle $\mu$ (Fig.~4),
\begin{equation}
	\tan{(\mu)} = { \left( {B_{\varphi} \over B_p} \right) }
	            = { \left( { \varphi r \over L } \right) } \ ,
\end{equation}
where $r$ is the radial distance from the loop symmetry axis,
$L$ is the loop length, and $\varphi$ is the rotation angle of the
twisted flux tube. The rotation angle $\varphi$ is simply defined by the
number $N_{turn}$ of turns, 
\begin{equation}
	\varphi = 2 \pi N_{turn} \ .
\end{equation}	
The free energy density $\varepsilon_{free} = \varepsilon_{\varphi}$, 
which is the difference between
the non-potential $\varepsilon_{np}=B_{np}^2/8\pi$ and the potential energy 
$\varepsilon_p=B_p^2/8\pi$,
where $B_{np}$ is the nonpotential field, and $B_p$ is the potential field 
(Aschwanden 2013b),
\begin{equation}
	\varepsilon_{free} = 
	(\varepsilon_{np} - \varepsilon_p) 
	= \varepsilon_{\varphi} = { B_{\varphi}^2 \over 8 \pi } \ ,
\end{equation}
can be expressed with Eqs.~(1) and (3) as a function of the rotation angle
$\varphi$, which shows that the free energy density is zero for no twist 
or for a 
potential field, but is monotonically increasing with the square of the
rotation angle $\varphi$ (using Eqs.~1 and 3), 
\begin{equation}
	\varepsilon_{free} 
	= {1 \over 8\pi} \left({B_p \varphi\ r \over L} \right)^2 \ ,
\end{equation}

The twisted helical field line shown in Fig.~3 clearly illustrates the dynamics 
of Parker's braiding model. If we continue to twist the helical
cylinder, say by one full turn, the resulting helical loop will have a
full turn, which we can continue until we obtain a ``slinky'' with
an arbitrary large number of turns, and hence, arbitrary large
free energy. The free energy density increases quadratically with the rotation angle,
$\varepsilon_{free} \propto \varphi^2$ (Eq.~4).
Mathematically, there is no upper limit on the number
of twist and on the free energy density, which illustrates Parker's dilemma
that braiding leads to infinite energies, if it is not reduced by occasional
magnetic reconnection episodes, or by nanoflare events above some critical
threshold of the rotation (or twist) angle $\varphi$. 

Let us now consider a slightly different model that we call the
{\sl minimum energy} model for short, referring to the minimum of
the free magnetic energy density. We start with the same configuration
of two parallel field lines (Fig.~3f), of which one is increasingly
twisted by a rotating footpoint motion (Fig.~3g). For a small twist
angle (say $\varphi=\pi/2$) the twisted field line has the same force-free
solution (Fig.~3g) as in Parker's braiding model (Fig.~3b). This
is the case until we reach a twist angle of a half turn ($\varphi=\pi$)
(Figs.~3c and Fig.~3h), at which point we encounter a bifurcation
in the path of possible force-free magnetic field solutions.
In Parker's scenario, the number of helical turns increases, while
the number of helical turns decreases in the minimum energy
model, until it reaches the potential field solution of no twist 
after a rotation angle of one full turn ($\varphi = 2\pi$) (Fig.~3j).
The different behavior results from the ambiguous path after the
bifurcation at $\varphi(t)=\pi$. The twisting process has 
two options, either to increase the rotation angle, or to
decrease the rotation angle. The minimum energy principle
requires the path of lower free energy density, because the lower energy state
has a higher statistical probability than the higher energy state.
Thus, the minimum energy principle sets an upper limit for the rotation angle,
\begin{equation}  
	| \varphi | \le \pi \ .
\end{equation}
The resulting dependence of the free energy density 
$\varepsilon_{\varphi}(\varphi)$ is 
shown in Fig.~5, for Parker's braiding scenario (with quadratically
increasing free energy as a function of time), while the minimum energy 
scenario follows a decreasing free energy density after passing the maximum 
rotation angle limit $\varphi(t) = \pi$ at the bifurcation point.

We can generalize our argument of the circular motion of a helically
twisted field line to random walk motion of braided field lines.
We simulate a random walk of the rotation angle $\varphi(t)$ 
that follows a diffusive pattern $\varphi(t) \propto t^{1/2}$
(Fig.~6, solid curve) for Parker's braiding scenario.
For the minimum energy scenario, however, a bifurcation sets in
at $|\varphi| \le \pi$ and the path evolves along smaller
rotation angles (Fig.~6, dashed curve), preventing divergence
to large free energies, large rotation angles $\varphi(t)$, or 
large misalignment angles $\mu(t)$, according to Eq.~(1).

\subsection{Random Footpoint Motions}

We now generalize the above argument to random walk of the photospheric 
footpoints analytically. A magnetic flux bundle with a square cross-section 
of width $w$ is considered. The magnetic perturbations can be described 
in terms of the shapes of the field lines. These shapes are given by 
functions $X (x_0, y_0, z,t)$ and $Y (x_0, y_0, z,t)$, where $x_0$ and 
$y_0$ are labels of the field lines. Each field line is given by a curve 
with cartesian coordinates $[ X (x_0, y_0, z,t) , Y (x_0, y_0, z,t), z]$, 
where $z$ parameterizes the curve. The three components of the magnetic 
field are
\begin{equation}
B_x = \frac{B_0}{J} \frac{\partial X}{\partial z} , ~~~~~ B_y = \frac{B_0}{J} \frac{\partial Y}{\partial z} , ~~~~~ B_z = \frac{B_0}{J} ,
\end{equation}
where $B_0$ is the strength of the original uniform field, and $J$ is the Jacobian on the mapping from $(x_0,y_0)$ to $(X,Y)$:
\begin{equation}
J(x_0, y_0, z,t) \equiv \frac{\partial X}{\partial x_0} \frac{\partial Y}{\partial y_0} - \frac{\partial Y}{\partial x_0} \frac{\partial X}{\partial y_0} .
\end{equation}
We assume the loop is much longer than its width ($L \gg w$), so that the parallel components of the field is nearly constant, $J = 1$. During the time interval before onset of reconnection ($0 < t < t_R$), the coordinates $(x_0, y_0)$ are given by the initial positions of the field lines, but later on some field lines will have reconnected and $(x_0,y_0)$ are no longer given by the initial coordinates. The energy of the system is given by a volume integral of the magnetic free energy density:
\begin{equation}
E (t) = \frac{B_0^2}{8 \pi} \int_0^L \int_0^w \int_0^w \left[ \left( \frac{\partial X} {\partial z} \right)^2 +
  \left( \frac{\partial Y} {\partial z} \right)^2 \right] dx_0 dy_0 dz .
\end{equation}
The boundary conditions for $X$ and $Y$ are the positions of the field lines at the boundary plates. A lower bound on the energy can be obtained by assuming that the field lines are nearly straight lines connecting the footpoints at $z = 0$ and $z = L$:
\begin{equation}
\frac{\partial X} {\partial z} \approx \frac{\Delta X}{L}  , ~~~~~~~ \frac{\partial Y} {\partial z} \approx \frac{\Delta Y}{L} ,
\end{equation}
where $\Delta X$ and $\Delta Y$ are relative displacements of the footpoints:
\begin{eqnarray}
\Delta X (x_0,y_0,t) & \equiv & X (x_0,y_0,L,t) - X (x_0,y_0,0,t) , \\
\Delta Y (x_0,y_0,t) & \equiv & Y (x_0,y_0,L,t) - Y (x_0,y_0,0,t) .
\end{eqnarray}
Then the minimum energy is
\begin{equation}
E_{\rm min} (t) = \frac{B_0^2}{8 \pi L} \int_0^w \int_0^w \left[ (\Delta X )^2 + (\Delta Y)^2 \right] dx_0 dy_0 .
\end{equation}
When the displacements $\Delta X$ and $\Delta Y$ become much larger than the correlation length $\ell$ of the footpoint motions, the field lines are forced to bend around each other, causing the formation of thin current sheets where reconnection can take place. The newly reconnected field lines will tend to be more straight and have smaller footpoint displacements, causing the magnetic stress to be reduced. Therefore, we expect that the footpoint displacements remain limited in magnitude,
\begin{equation}
| \Delta X | \sim | \Delta Y | \sim \ell ,
\end{equation}
and the minimum energy will saturates at a value given by
\begin{equation}
E_{\rm min} \sim \frac{B_0^2 w^2 \ell^2}{4 \pi L} .
\end{equation}
Note that this expression (Eq.~14) of the free energy calculated based
on random motion of the footpoints is equivalent to the free energy 
density $\varepsilon_{free}$ 
defined for a helically twisted loop (Eq.~4), if we define the
2-D correlation length with $\ell = \sqrt{2} r \varphi$ and execute
the volume integral $E_{\rm min} = \varepsilon_{free} \ V$ of the
free energy density $\varepsilon_{free}$ with the volume definition 
$V = w^2 L$, based on the width $w$ and length $L$ of a loop. 

We suggest the actual free energy $E$ will not be much larger 
than $E_{\rm min}$. The above analysis is consistent with the 
work of Rappazzo \& Parker (2013) and Rappazzo (2015), who argue 
that there is a threshold for the formation of thin current sheets, 
$B_\perp \sim B_0 \ell / L$, which corresponds to Eq.~(14). 
Therefore, the minimum energy scenario suggests that magnetic 
reconnection will prevent the build-up of arbitrary large transverse 
fields. The transverse displacements remain on the order of the scale 
of the solar granulation, $\ell \approx 1000$ km, and a strongly braided 
field does not form.

\section{	DISCUSSION 		}

\subsection{	Criticism of Parker's Braiding Scenario		}

The ``magnetic field braiding'' scenario of Parker (1988) suggests
that {\it the X-ray corona is created by the dissipation of the many
tangential discontinuities arising spontaneously in the bipolar
fields of the active regions of the Sun as a consequence of random
continuous motion of the footpoints of the field in the photospheric
convection.} This concept implies that the field lines become
increasingly more twisted and braided by the random motion
of the footpoints, as depicted in the cartoons of Parker's (1983) 
scenario (Fig.~1). Although cartoons are not intended to provide
a scientific accurate model on scale, they should convey 
the correct concept qualitatively. However, three major criticisms  
can be discussed about the cartoons featured in Fig.~1: (i) the
discontinuity of the magnetic field, (ii) the large misalignment
angles, and (iii) the location of nanoflares. 

The first criticism, i.e., the discontinuity of the magnetic field
between a braided field line and a unbraided (or less-braided) field
line is most conspicuously depicted in Fig.~1d. Even if both types
of field lines would correspond to a physical solution of Maxwell's
or the ideal MHD equations, they cannot evolve independently of each 
other, as the cartoon in Fig.~1d suggests. If one field line becomes 
twisted, a continuous field solution (that fulfills the 
divergence-freeness and force-freeness) affects the adjacent field 
lines in such a way that only small misalignment angles between two 
neighbored field lines occur, as depicted in Fig.~2. In this sense, 
discontinuities in the 3-D magnetic field solution are unobserved
and unphysical.

The second criticism, i.e., the large misalignment angles between
adjacent magnetic field lines predicted by Parker's braiding model
is related to the unphysical discontinuity of the magnetic field.
We can estimate the typical misalignment angle between two adjacent
field lines in the following way. Two field lines with a distance
$r$ and $r+\Delta r$ from the symmetry axis of a flux tube have
an angle of $\mu$ and $\mu+\Delta \mu$ relative to the symmetry axis, and
the misalignment angle $\Delta \mu$ between the two adjacent field
lines is according to Eq.~(1),
\begin{equation}
	\Delta \mu = \arctan{ \left( {\varphi [r+\Delta r] \over L}\right)} - 
	             \arctan{ \left( {\varphi r \over L }\right)} \ ,
\end{equation}
For instance, for field lines with distances of $r=0,1,2,...,10$ Mm
from the flux tube axis, at a relative distance of $\Delta r=1$ Mm,
a length of $L=100$ Mm, and twisted by a half turn ($\phi=\pi$),
we obtain with Eq.~(6) misalignment angles of
$\Delta \mu \approx 1.6^\circ-1.8^\circ$. Such scales of small
misalignment angles are approximately rendered in the revised
cartoons shown in Fig.~2, in contrast to the large misalignment
angles of order $\Delta \mu \approx 45^\circ$ shown in Fig.~1. 
Also, observations with TRACE and AIA/SDO of the corona in EUV 
wavelengths show invariably near-parallel loops, e.g., as
determined from stereoscopic triangulation (Aschwanden et al.~2008).

The third criticism of Parker's braiding scenario is the location
of nanoflares, which would be expected throughout the corona, since
uniformly twisted flux tubes have a constant twist angle
(or a constant $\alpha$-parameter) along the loops. In contrast,
if a divergence-free and force-free solution of magnetic fields
is calculated (such as depicted in Fig.~2), misalignment angles
in the force-free corona are small, and therefore no tangential
discontinuities arise, and thus no nanoflares are produced in the
force-free corona. However, the chromosphere and parts of the
transition region are not force-free, and thus tangential
discontinuities and nanoflaring are expected there. This is
indicated with misalignments of the chromospheric footpoints
in Fig.~2. According to measurements and modeling of the 3-D 
magnetic field with the virial theorem, the chromosphere was
found to be not force-free up to a height of $h \approx 400$ km
(Metcalf et al.~1995). Another (observational) argument 
for nanoflare locations 
in the chromosphere and transition region are EUV observations
(Aschwanden et al.~2000) and hard X-ray observations of microflares
(Hannah et al.~2001).

Of course, these three criticisms, (i) the discontinuity of the field, 
(2) the misalignment of adjacent field lines, and (iii) the
location of nanoflares are all related to each other. If there
is no discontinuity, then there is no misalignment, and no nanoflaring,
but the three aspects can be measured from observations each separately.

\subsection{	Limits of the Energy Build-Up 			}

A new aspect of this critics on Parker's nanoflare scenario is the
minimum energy aspect. Using the concept of helical twisting in the
build-up of free energy we have shown that the free energy evolves
in a nonlinear (quadratic) way with the twisting or rotation angle,
i.e., $\varepsilon_{\varphi} \propto \varphi^2$, which introduces a bifurcation
at $|\varphi(t)|=\pi$. Whenever such a bifurcation point is reached
during the random shuffling of the footpoints of coronal loops, 
the lower energy state is statistically more likely to be chosen,
which prevents an infinite build-up of free energy.

This is also consistent with the kink instability criterion 
(T\"or\"ok and Kliem 2003; Kliem et al.~2004), 
which limits the maximum stable solution to 
about $\lapprox 1$ helical turn (Hood and Priest 1979; Sakurai 1976;
Mikic et al.~1990). If loops or filaments are helically twisted by a larger
number or turns, they become unstable and erupt.

\subsection{	Numerical MHD Simulations			}

Many authors have studied magnetic braiding using 3-D MHD simulations 
(e.g., Mikic et al.~1989; Hendrix et al.~1996; Longcope and Sudan 1994;
Ng and Bhattacharjee 2008; Galsgaard and Nordlund 1996, 1999;
Rappazzo 2007, 2008; Dahlburg et al.~2016). This work has provided much 
insight into the structure and topology of braided fields, and the 
formation of current sheets in such models. However, none of these models 
has been able to reproduce the strongly braided fields predicted by 
Parker (1983). In most cases the field lines deviate only slightly 
from straight lines connecting the footpoints at the two boundary plates, 
consistent with the predictions of the minimum energy scenario.

In most cases detailed comparisons between observed and predicted coronal 
heating rates have not been made. Therefore, it is still an open question 
whether magnetic braiding models based on results from 3-D MHD simulations 
(as opposed to cartoons) are consistent with the observed properties of 
coronal loops. In order to have a valid comparison, the footpoint motions 
imposed in magnetic braiding models must be consistent with the observed 
motions of magnetic elements on the photosphere. In particular, 
observations of random-walk diffusion constants (see Berger 1998, and 
references therein) provide important constraints on the rate at which 
magnetic energy can be injected into coronal loops by random footpoint 
motions. The imposed footpoint motions should be consistent with these 
diffusion constants, otherwise the coronal heating rate may be overestimated.

The 3-D MHD model for a coronal loop has been extended to include an 
approximate description of the lower atmosphere 
(van Ballegooijen et al.~2011, 2014; Asgari et al.~2012, 2013, 2014, 2015). 
In this model Alfv\'{e}n waves are launched in kilogauss flux tubes in the 
photosphere, and the waves travel upward into the corona, where wave 
dissipation and plasma heating take place. Magnetic braiding occurs in 
these models, but the braiding is more dynamic in nature because the 
Alfv\'{e}n travel time from one photosphere footpoint to the other 
(about 2 minutes) is comparable to the timescale of the imposed 
footpoint motions. Therefore, when the lower atmosphere is included in 
the model, the corona responds much more dynamically to the footpoint 
motions, not quasi-statically as assumed in the Parker's braiding model 
and in the cascade model (van Ballegooijen 1986). Quasi-static evolution 
was recovered only when the lower atmosphere was removed from the model 
(see section 3 in van Ballegooijen et al.~2014). This dynamic behavior 
for models that include the lower atmosphere is a consequence of the much 
higher density of these layers compared to the corona.

An observed coronal loop is expected to be rooted in multiple photospheric 
flux elements, and in plage regions these elements are expected to merge 
into a space-filling field at some height in the chromosphere. Recently, 
van Ballegooijen et al.~(2017) simulated the dynamics of Alfv\'{e}n waves 
in a coronal loop rooted in multiple flux tubes. They also developed a 
second ``magnetic braiding" model in which the footpoint motions are 
applied in the low chromosphere, and realistic photospheric velocities 
and correlation times are used (see section 5 of that paper). 
They found that the energy injected into the corona by magnetic braiding 
is much less than that provided by Alfv\'{e}n waves, and is insufficient 
to compensate for the radiative and conductive losses from the corona. 
The main reason for this low energy input rate is that the energy builds 
up only for about 20 minutes, after which the energy dissipation rate 
becomes equal to the energy input rate. This 20-minute build-up time is 
much shorter than the 8-hour period shown to be required by Parker (1983). 
Therefore, a strongly braided field never develops in this model.

The above result is consistent with predictions from the cascade model 
(van Ballegooijen 1986). According to this model, quasi-static braiding 
causes the thickness $\delta (t)$ of the coronal current sheets to decrease 
exponentially with time, $\delta (t) \sim e^{-t/\tau}$. The time scale 
$\tau$ is determined by the dynamical time of the footpoint motions and 
is only a few minutes. When $\delta (t)$ reaches the magnetic diffusion 
scale, reconnection will start and will soon dominate the evolution of 
the magnetic structure. The onset of reconnection occurs at a time 
$t_R \approx \onehalf \tau \ln R_m$, where $R_m$ is the magnetic Reynolds 
number. Even for the very high Reynolds numbers found on the Sun 
($R_m \sim 10^{10}$), we expect $t_R < 12 \tau < 30$ minutes, much less 
than the 8-hour period obtained by Parker (1983). In numerical models 
$R_m$ is much smaller and $t_R$ is somewhat reduced, but we do not expect 
that $t_R$ can be significantly increased simply by increasing the spatial 
resolution of the numerical models (with higher resolution resistive 
instabilities should develop as well). Therefore, the absence of a 
strongly braided field in the second simulation by van Ballegooijen 
et al.~(2017) is a real effect and is not due to a lack of spatial 
resolution in the simulation.

\subsection{Consequences for Coronal Heating}

MHD simulations of coronal loops indicate that quasi-static magnetic 
braiding on the scale of the solar granulation ($\ell \sim 1000$ km) 
cannot provide enough energy to heat active region cores to the observed 
high temperatures and densities. Therefore, we must look for other 
sources of energy. One possibility is Alfv\'{e}n waves, which are 
believed to be important for heating the plasma in coronal holes 
and driving the solar wind, but may also be important for coronal 
loops (e.g., Antolin and Shibata 2010; van Ballegooijen et al.~2011). 
In recent modeling (van Ballegooijen et al.~2017) it was found that 
Alfv\'{e}n wave heating can produce a loop with a peak temperature 
of 2.5 MK and pressure of about 1.8 $\rm dyne ~ cm^{-2}$. However, 
the model cannot fully explain the observed differential emission 
measure (DEM) distributions (Warren et al.~2012; Schmelz et al.~2015), 
which have a peak at about 4 MK and extend to both lower and higher 
temperatures.

The non-existence of highly twisted coronal loops eliminates coronal 
nanoflares as a possible energy source. This is not at odds with 
observations, because no convincing direct observational evidence 
of coronal nanoflares has been put forward so far. Searches for the 
coronal type of nanoflares as predicted by Parker (1983) have been 
unsuccessful. In contrast, there exists a type of observed EUV 
nanoflares, which have typical energies of $E \approx 10^{24}$ erg, 
but occur in the lowest layers of the solar corona, near the 
transition region (Krucker \& Benz 1998; Parnell \& Jupp 2000; 
Aschwanden et al.~2000). Such events may produce energetic 
electrons that travel upward along the loop and deposit their 
energy near the loop top (in form of {\sl direct heating}),
or may accelerate electrons that travel downward along the loop
and deposit their energy in the chromosphere as it occurs
in the thick-target model of normal flares. 
The upward directed nonthermal energy flux would 
contribute to the direct heating of the coronal plasma, but unlike 
braided fields or MHD waves, would not be detectable
in an unambiguous way.

Potential tracers of coronal nanoflares in the definition of 
Parker have been hypothesized in the form of a coronal 
high-temperature component, $T > 8$ MK (Reale et al. 2009; 
Schmelz et al. 2009), but evidence is scant and dubious, 
considering the uncertainties of DEM modeling at high 
temperatures of $T_e > 10$ MK. The original motivation of 
Parker's nanoflare scenario has been the finding of an 
ubiquitous mechanism that can transport magnetic energy 
from the sub-photospheric convection zone to the
corona and to dissipate it there. However, based on the 
lack of observational evidence for misaligned loops, 
the lack of coronal nanoflares, and the plausibility of the
minimum energy principle, we suggest to revise the scenario 
of the Parker-type coronal nanoflares by attributing the 
misaligned magnetic fields and the generation of nanoflares 
to the lowest parts of the solar atmosphere (near the 
chromosphere and transition region), rather than to the 
upper corona as suggested by Parker (1983).

\section{	CONCLUSIONS 		}

We revisit Parker's coronal loop braiding model in the light
of a new analytical magnetic field model that can calculate 
approximate solutions of a divergence-free and force-free
magnetic field, based on helically twisted loop structures.
This analytical code allows us also to study the coronal 
braiding and nanoflaring scenario of Parker (1983, 1988).
We arrive at the following conclusions:

\begin{enumerate}

\item{Simplifying the random walk trajectory of a coronal
footpoint motion to a circular path (without loss of 
generality in Parker's model), 
the force-free magnetic field solution is a helically twisted loop 
whose footpoint is rotated with a time-dependent rotation angle
$\varphi(t)$, and the free energy increases quadratically
with the rotation angle, i.e., $\varepsilon_{free}(t) \propto \varphi^2(t)$,
which implies that the free (magnetic) energy can grow to arbitrary
large values in Parker's scenario.} 

\item{We employ the minimum energy principle to the footpoint
motion of helically twisted loops whenever there is a
bifurcation between multiple force-free field solutions.
This obeys the principle of the highest statistical likelihood,  
prevents the build-up of infinite energy, and yields small
misalignment angles between adjacent loops, without creating
tangential discontinuities or producing nanoflares in the corona.} 

\item{We estimate the misalignment angles between two adjacent
loops (separated by $\Delta r = 1$ Mm, for separation distances
of $r=1-10$ Mm, and with twisting by a half turn) and obtain
relatively small misalignment angles of $\Delta \mu = 
1.6^\circ-1.8^\circ$ between adjacent loops,
which confirm the observations of
ubiquitous near-parallel loops seen in EUV. The Parker model
predicts substantially larger misalignment angles  
and a threshold value of $\mu_{crit} \ge 20^\circ$ 
is required for nanoflaring (Parker 1983).}

\item{The Parker scenario predicts that the location of nanoflares
is distributed throughout the corona, because individual field lines
or flux tubes are uniformly twisted along their length (since
the non-potential $\alpha$-parameter is constant along each field
line). This spatial prediction of nanoflares in the entire corona
is not consistent with observations, because all small EUV 
nanoflares and hard X-ray microflares are found to be localized
in the lowest part of the solar atmosphere. Therefore, nanoflares
are more likely to occur in the lower atmosphere (in the chromosphere
and transition region), where the magnetic field is not force-free.}
\end{enumerate}

In summary, while Parker's braiding model faces the three
major problems of: (i) the (unphysical and unobserved) 
discontinuity of the magnetic field,
(ii) large (unobserved) misalignment angles, and (iii) the
(unobserved) location of coronal nanoflares. In contrast,
the minimum energy model of helically twisted loops offers:
(i) a continuous 3-D magnetic field solution without discontinuities,
(ii) small misalignment angles ($\Delta \mu \lapprox 1.6^\circ-1.8^\circ $) that are
consistent with observations, and (iii) nanoflare locations in the
lower atmosphere (chromosphere to transition region) where 
EUV nanoflares and hard X-ray microflares are observed indeed.
In short, all problems of Parker's braiding and nanoflaring model 
can be reconciled with the minimum energy principle.

\section*{APPENDIX A: VERTICAL-CURRENT APPROXIMATION NONLINEAR
FORCE-FREE FIELD MODEL}

A physically valid coronal magnetic field solution has to satisfy
Maxwell's equations, which includes the divergence-freeness condition,
$$
	\nabla \cdot {\bf B} = 0 \ ,
	\eqno(A1)
$$
and the force-freeness condition,
$$
	\nabla \times {\bf B} = \alpha({\bf r}) {\bf B} \ ,
	\eqno(A2)
$$
where $\alpha({\bf r})$ represents a scalar function that depends
on the position ${\bf r}$, but is constant along a magnetic field
line. Three different types of magnetic fields are generally
considered for applications to the solar corona: (i) a {\sl potential
field (PF)} where the $\alpha$-parameter vanishes $(\alpha=0)$,
(ii) a {\sl linear force-free field (LFFF)} $(\alpha = const)$, and
(iii) a {\sl nonlinear force-free field (NLFFF)} with a spatially
varying $\alpha({\bf r}) \neq 0$.

Due to the nonlinearity of the equation system, no general
analytical solution of the magnetic field ${\bf B}({\bf r})$ has
been obtained for the coupled equation system of (A1)-(A2).
However, an analytical approximation of a divergence-free and
force-free magnetic field solution has been derived for a vertical
current at the lower photospheric boundary, which twists a
field line into a helical shape (Aschwanden 2013a),
$$
        B_r(r, \theta) = B_0 \left({d^2 \over r^2}\right)
        {1 \over (1 + b^2 r^2 \sin^2{\theta})} \ ,
	\eqno(A3)
$$
$$
        B_\varphi(r, \theta) =
        B_0 \left({d^2 \over r^2}\right)
        {b r \sin{\theta} \over (1 + b^2 r^2 \sin^2{\theta})} \ ,
	\eqno(A4)
$$
$$
        B_\theta(r, \theta) \approx 0 \ ,
	\eqno(A5)
$$
$$
        \alpha(r, \theta) \approx {2 b \cos{\theta} \over
        (1 + b^2 r^2 \sin^2{\theta})}  \ ,
	\eqno(A6)
$$
where ${\bf B}({\bf r}) = [B_r, B_{\varphi}, B_{\theta}]$ is the
magnetic field in a spherical coordinate system $[r, \varphi, \theta]$,
$B_0$ is the magnetic field strength vertically above a buried magnetic
charge at photospheric height $h=0$, $d$ is the depth of the buried
magnetic charge, and $b$ is a parameter related to the nonlinear
$\alpha$-parameter. We see that the non-potential solution $(b \neq 0$)
degenerates to the potential field solution in the case of $(b=0)$,
$$
	B_r(r) = B_0 \left( {d_j \over r_j} \right)^{2} \ .
	\eqno(A7)
$$
The 3-D vector field of the magnetic field is then,
$$
        {\bf B_j}({\bf x})
        = B_j \left({d_j \over r_j}\right)^2 {{\bf r}_j \over r_j} \ ,
	\eqno(A8)
$$
Such a magnetic field model with a single buried magnetic (unipolar)
charge can be adequate for a sunspot. For a bipolar active region,
at least two magnetic charges are necessary.

A general magnetic field can be constructed by superposing the
$N_m$ fields of $j=1,...,N_m$ magnetic charges, defined as,
$$
        {\bf B}({\bf x}) = \sum_{j=1}^{N_{\rm m}} {\bf B}_j({\bf x})
        = \sum_{j=1}^{N_{\rm m}}  B_j
        \left({d_j \over r_j}\right)^2 {{\bf r_j} \over r_j} \ .
	\eqno(A9)
$$
where the depth $d_j$ of a magnetic charge $j$ is,
$$
	d_j = 1-\sqrt{x_j^2+y_j^2+z_j^2} \ ,
	\eqno(A10)
$$
and the distance $r_j$ between the magnetic charge position $(x_j, y_y, z_j)$
and an arbitrary location $(x,y,z)$ where the calculation of a magnetic field
vector is desired, is defined by,
$$
 	r_j = \sqrt{(x-x_j)^2+(y-y_j)^2+(z-z_j)^2} \ .
	\eqno(A11)
$$
The magnetic field ${\bf B}(x,y,z)$ in Cartesian coordinates can be
transformed into spherical coordinates ${\bf B}({\bf r}) = 
[B_r, B_{\varphi}, B_{\theta}]$, as expressed in Eqs.~(A3)-(A6).

The multi-pole magnetic field is divergence-free, since
$$
        \nabla \cdot {\bf B} = \nabla \cdot (\sum_j {\bf B}_j)
        = \sum_j (\nabla \cdot {\bf B}_j) = 0 \ ,
	\eqno(A12)
$$
while the force-freeness is fulfilled with second-order accuracy
in $\alpha^2 \propto [ b r \sin{ \theta } ]^2$ for the solution
of the vertical-current approximation of Eqs.~(A3)-(A6) 
(Aschwanden 2013a),
$$
        \nabla \times {\bf B} =
        \nabla \times \sum_{j=1}^{N_{\rm m}} {\bf B}_j =
        \sum_{j=1}^{N_{\rm m}} (\nabla_j \times {\bf B}_j) =
        \sum_{j=1}^{N_{\rm m}} \alpha_j ({\bf r}) {\bf B}_j =
        \alpha ({\bf r}) {\bf B} \ .
	\eqno(A13)
$$
Numerical tests of comparing the analytical approximation solution
with other nonlinear force-free field codes have been conducted for
a large number of simulated and observed magnetograms, and
satisfactory agreement with other NLFFF codes has been established
(e.g., Aschwanden 2013a, 2013b, 2016; Aschwanden and Malanuchenko
2013; Warren et al.~2018).
	
\acknowledgements

We acknowledge useful discussions with Robertus Erdelyi, 
Ma Asgari-Targhi, and other attendees of a DKIST science planning
meeting in April 2018, held at Newcastle upon Thyne (UK).
This work was partially supported by NASA contract NNX11A099G
``Self-organized criticality in solar physics'', NASA contract
NNG04EA00C of the SDO/AIA instrument, and NASA contract
NNG09FA40C of the IRIS mission.

\clearpage

\section*{	REFERENCES	}

\def\ref#1{\par\noindent\hangindent1cm {#1}}

\ref{Antolin, P. and Shibata, K. 2010, ApJ 712, 494.
 	{\sl The Role of Torsional Alfv\'en Waves in Coronal Heating}}
\ref{Aschwanden, M.J., Tarbell, T., Nightingale, R., et al. 2000, ApJ 535, 1047.
        {\sl Time variability of the quiet Sun observed with TRACE.
        II. Physical parameters, temperature evolution, and energetics
        of EUV nanoflares}}
\ref{Aschwanden, M.J. 2005,
	{\sl Physics of the Solar Corona}, Springer: Berlin.}
\ref{Aschwanden, M.J., Wuelser, J.P., Nitta, N., and Lemen,J.
 	2008, ApJ 679, 827.
 	{\sl First 3D reconstructions of coronal loops with the STEREO A and B 
	spacecraft: I. Geometry}}
\ref{Aschwanden, M.J. 2013a, SoPh 287, 323.
	{\sl A nonlinear force-free magnetic field approximation
	suitable for fast forward-fitting to coronal loops. I. Theory}}
\ref{Aschwanden, M.J. 2013b, SoPh 287, 369.
	{\sl A nonlinear force-free magnetic field approximation
	suitable for fast forward-fitting to coronal loops. 
	III. The free energy}}
\ref{Aschwanden, M.J. and Malanuchenko, A. 2013, SoPh 287, 345.
	{\sl A nonlinear force-free magnetic field approximation
	suitable for fast forward-fitting to coronal loops. 
	II. Numerical Code and Tests}}
\ref{Aschwanden, M.J. 2016, ApJSS 224, 25.
	{\sl The vertical current approximation nonlinear force-free
	field code - Description, performance tests, and measurements
	of magnetic energies dissipated in solar flares}}
\ref{Asgari-Targhi,M., and van Ballegooijen, A.A. 2012,
 	ApJ 746, 81.
 	{\sl Model for Alfv\'en Wave Turbulence in Solar Coronal Loops: 
	Heating Rate Profiles and Temperature Fluctuations}}
\ref{Asgari-Targhi,M., van Ballegooijen, A.A., Cranmer, S.R., and DeLuca, E.E.
 	2013, ApJ, 773, 111.
 	{\sl The Spatial and Temporal Dependence of Coronal Heating by 
	Alfv\'en Wave Turbulence}}
\ref{Asgari-Targhi,M., van Ballegooijen, A.A., and Imada, S. 2014,
 	ApJ 786, 28.
 	{\sl Comparison of Extreme Ultraviolet Imaging Spectrometer 
	Observations of Solar Coronal Loops with Alfv\'en Wave Turbulence 
	Models}}
\ref{Asgari-Targhi, M., Schmelz, J.T., Imada, S., Pathak, S., and
	Christian, G.M. 2015, ApJ 807, 146.
 	{\sl Modeling of Hot Plasma in the Solar Active Region Core}}
\ref{Berger, T.E., Loefdahl, M.G., Shine, R.A., and Title, A.M. 1998,
 	ApJ 506, 439. 
 	{\sl Measurements of solar magnetic element dispersal}}
\ref{Cargill, P.J., Warren, H.P., and Bradshaw, S.J. 2015,
 	Royal Society of London Philosophical Transactions Series A, 373, 40260.
 	{\sl Modelling nanoflares in active regions and implications for 
	coronal heating mechanisms}}
\ref{Dahlburg, R.B., Einaudi, G., Taylor, B.D., Ugarte-Urra, I., 
	Warren, H.P., Rappazzo, A.F., and Velli, M. 2016,
 	ApJ 817, 47.
 	{\sl Observational Signatures of Coronal Loop Heating and Cooling 
	Driven by Footpoint Shuffling}}
\ref{Galsgaard, K. and Nordlund, A. 1996, JGR 101/A6, 13445.
 	{\sl Heating and activity of the solar corona. 1. Boundary shearing 
	of an initially homogeneous magnetic field}}
\ref{Galsgaard, K. and Nordlund, A. 1997, JGR 102, 219.
 	{\sl Heating and activity of the solar corona. 2. Kink instability 
	in a flux tube}}
\ref{Hannah,I.G., Hudson, H. S., Battaglia, M., Christe, S., 
	Kasparova, J., Krucker, S., Kundu, M. R., and Veronig, A.
 	2011, SSRv 159, 263. 
 	{\sl Microflares and the Statistics of X-ray Flares}}
\ref{Hendrix, D.L., van Hoven, G., Mikic, Z., and Schnack, D.D. 1996,
 	ApJ 470, 1192.
 	{\sl The viability of ohmic dissipation as a coronal heating source}}
\ref{Hood, A.W. and Priest, E.R. 1979, A\&A 77, 233.
 	{\sl The equilibrium of solar coronal magnetic loops}}
\ref{Kliem, B., Titov, V.S., and Toeroek, T. 2004, A\&A 413, L23.
 	{\sl Formation of current sheets and sigmoidal structure 
	by the kink instability of a magnetic loop}}
\ref{Klimchuk, J.A. 2015,
 	Royal Society of London Philosophical Transactions Series A, 373, 40256.
 	{\sl Key aspects of coronal heating}}
\ref{Krucker, S. and Benz, A.O. 1998, ApJ 501, L213.
        {\sl Energy distribution of heating processes in the quiet
 	solar corona}}
\ref{Longcope,D.W. and Sudan,R.N. 1994, ApJ 437, 491.
 	{\sl Evolution and statistics of current sheets in coronal 
	magnetic loops}}  
\ref{Lorenz, E.N. 1963, J.Atmos Sci. 20, 130.
	{\sl Deterministic nonperiodic flow}}
\ref{Metcalf, T.R., Jiao, L., Uitenbroek,H., McClymont,A.N., and Canfield,R.C.
 	1995, ApJ 439, 474.
 	{\sl Is the solar chromospheric magnetic field force-free?}}
\ref{Mikic, Z., Schnack, D.D., and van Hoven, G. 1989,
 	ApJ 338, 1148.
 	{\sl Creation of current filaments in the solar corona}}
\ref{Mikic, Z., Schnack,D.D., and van Hoven,G. 1990, ApJ 361, 690.
 	{\sl Dynamical evolution of twisted magnetic flux tubes.
	I. Equilibrium and linear stability}}
\ref{Ng, C.S., and Bhattacharjee, A. 2008, ApJ 675, 899.
 	{\sl A Constrained Tectonics Model for Coronal Heating}}
\ref{Parker, E.N. 1972, ApJ 174, 499.
	{\sl Topological dissipation and the small-scale fields
	in turbulent gases}}
\ref{Parker, E.N. 1983, ApJ 264, 642.
	{\sl Magnetic neutral sheets in evolving fields.
	II. Formation of the solar corona}}
\ref{Parker, E.N. 1988, ApJ 330, 474.
	{\sl Nanoflares and the solar X-ray corona}}
\ref{Parker, E.N. 1989, SoPh 121, 271.
	{\sl Solar and stellar magnetic fields and atmospheric
	structures: Theory}}
\ref{Parnell, C.E. and Jupp, P.E. 2000, ApJ 529, 554.
        {\sl Statistical analysis of the energy distribution of nanoflares
        in the Quiet Sun}}
\ref{Priest, E.R. 1982,
 	Geophysics and Astrophysics Monographs Volume 21, 
	D.Reidel Publishing Company, Dordrecht {\sl Solar Magnetohyrdodynamics}}
\ref{Rappazzo, A.F., Velli, M., Einaudi, G., and Dahlburg, R.B. 2007,
 	ApJ 657, L47.
 	{\sl Coronal Heating, Weak MHD Turbulence and Scaling Laws}
\ref{Rappazzo, A.F. Velli, M., Einaudi, G., and Dahlburg, R.B. 2008,
 	ApJ 677, 1348.}
 	{\sl Nonlinear Dynamics of the Parker Scenario for Coronal Heating}}
\ref{Rappazzo, A.F. and Parker, E.N. 2013,
 	ApJ 773, L2.
 	{\sl Current Sheets Formation in Tangled Coronal Magnetic Fields}}
\ref{Rappazzo, A.F. 2015,
 	ApJ 815, 8,
 	{\sl Equilibria, dynamics, and current sheet formation in magnetically 
	confined coronae}}
\ref{Reale, F. McTiernan, J.M., and Testa, P. 2009, ApJ 704, 58.
        {\sl Comparison of Hinode/XRT and RHESSI detection of hot plasma
        in the non-flaring solar corona}}
\ref{Sakurai K. 1976, PASJ 28, 177.
	{\sl Magnetohydrodynamic interpretation of the motion of prominences}}
\ref{Schmelz, J.T., Kashyap, V.L., Saar, S.H., et al. 2009, ApJ 704, 863.
        {\sl Some like it hot: Coronal heating observations
        from Hinode X-ray Telescope and RHESSI}}
\ref{Schmelz, J.T., Asgari-Targhi, M., Christian, G.M., Dhaliwal, R.S., 
	and Pathak, S. 2015,
 	ApJ 806, 232.
 	{\sl Hot Plasma from Solar Active Region Cores: a Test of AC 
	and DC Coronal Heating Models?}}
\ref{Tenerani, A., Velli, M., Pucci, E., Landi, S., Rappazzo, A.F. 2016,
	J.Plasma Physics 82/5, 535820501.
	{\sl Ideally unstable current sheets and the triggering of
	fast magnetic reconnection}}
\ref{T\"or\"ok, T. and Kliem, B. 2003, A\&A 406, 1043.
	{\sl The evolution of twisting coronal magnetic flux tubes}}
\ref{van Ballegooijen, A.A. 1985, 
	ApJ 298, 421.
	{\sl Electric currents in the solar corona and the existence of
	magnetostatic equilibrium}}
\ref{van Ballegooijen, A.A. 1986, 
	ApJ 311, 1101.
	{\sl Cascade of magnetic energy as a mechanism of coronal heating}}
\ref{van Ballegooijen, A.A. 1988, Geophys.Astrophys.Fluid Dynamics
	41(3), 181.
	{\sl Force free fields and coronal heating part I.
	The formation of current sheets}}
\ref{van Ballegooijen, A.A., Asgari-Targhi, M., Cranmer, S.R., and DeLuca, E.E.
 	2011, ApJ 736, 3.
 	{\sl Heating of the Solar Chromosphere and Corona by Alfv\'en Wave 
	Turbulence}}
\ref{van Ballegooijen, A.A., Asgari-Targhi, M., and Berger, M.A. 2014,
	ApJ 787, 87.
	{\sl On the relationship between photospheric footpoint motions
	and coronal heating in solar active regions}}
\ref{van Ballegooijen, A. A., Asgari-Targhi, and Voss, A. 2017, 
	Apj 849, 46.
	{\sl The heating of solar coronal loops by Alfv\'en wave turbulence}}
\ref{Warren, H.P., Winebarger, A.R., and Brooks, D.H. 2012,
 	ApJ 759, 141.
 	{\sl A Systematic Survey of High-temperature Emission in Solar Active 
	Regions}}
\ref{Warren, H.P., Crump, N.A., Ugarte-Urra, I., Sun, X., 
	Aschwanden, M.J., and Wiegelmann, T. 2018, ApJ 860, 46.
	{\sl Toward a quantitative comparison of magnetic field
	extrapolations and observed coronal loops}}

\clearpage

\begin{figure}
\centerline{\includegraphics[width=1.0\textwidth]{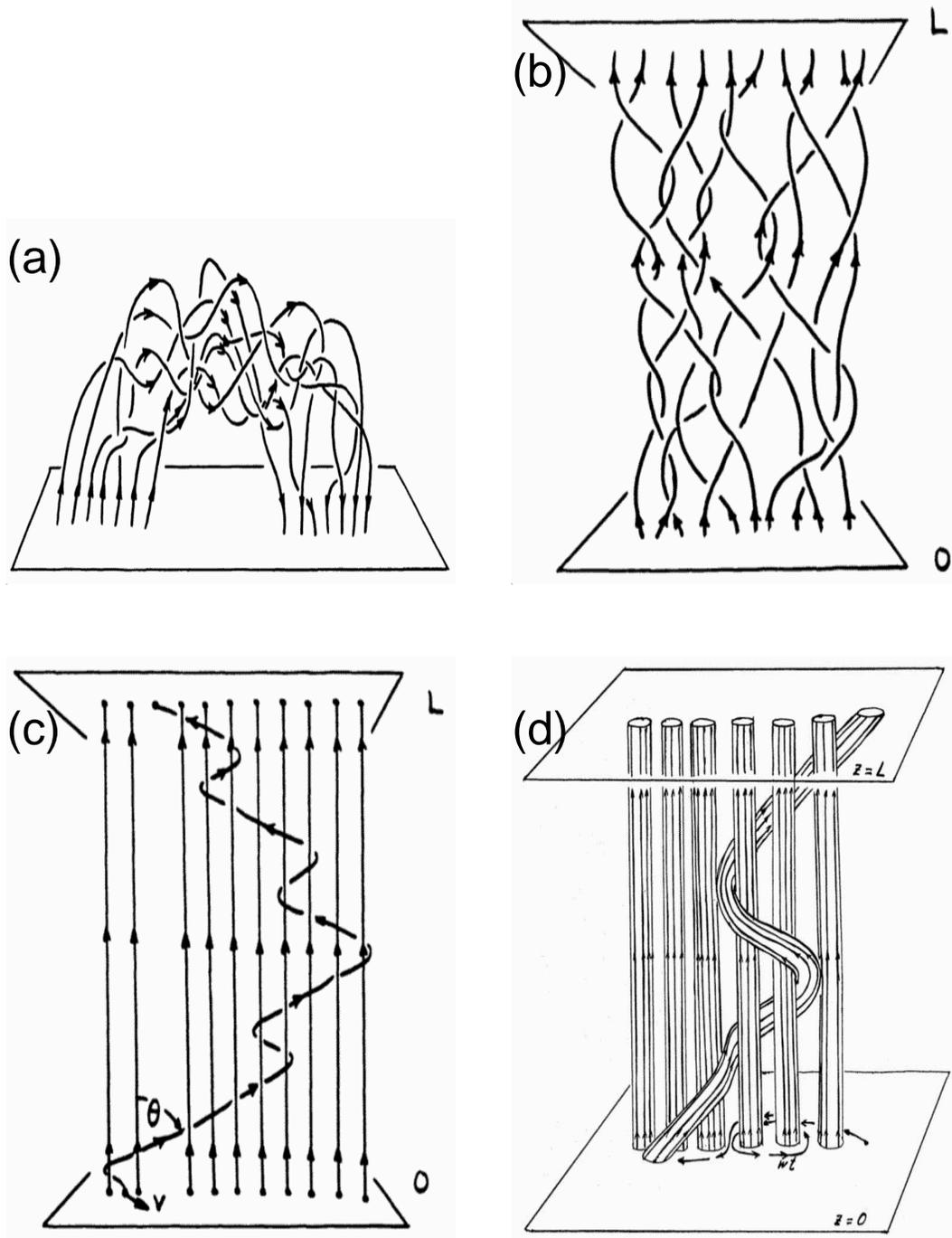}}
\caption{Original cartoons published by Eugene Parker (1989):
Schematic drawings of field lines of a bipolar magnetic reconnection
region above the photosphere (a); a bundle of loops stretched out in
vertical direction (b); a wandering (braiding) field line among
neighboring field lines (c), and a single flux tube after some
random walk of its footpoint (d).}
\end{figure}

\begin{figure}
\centerline{\includegraphics[width=1.0\textwidth]{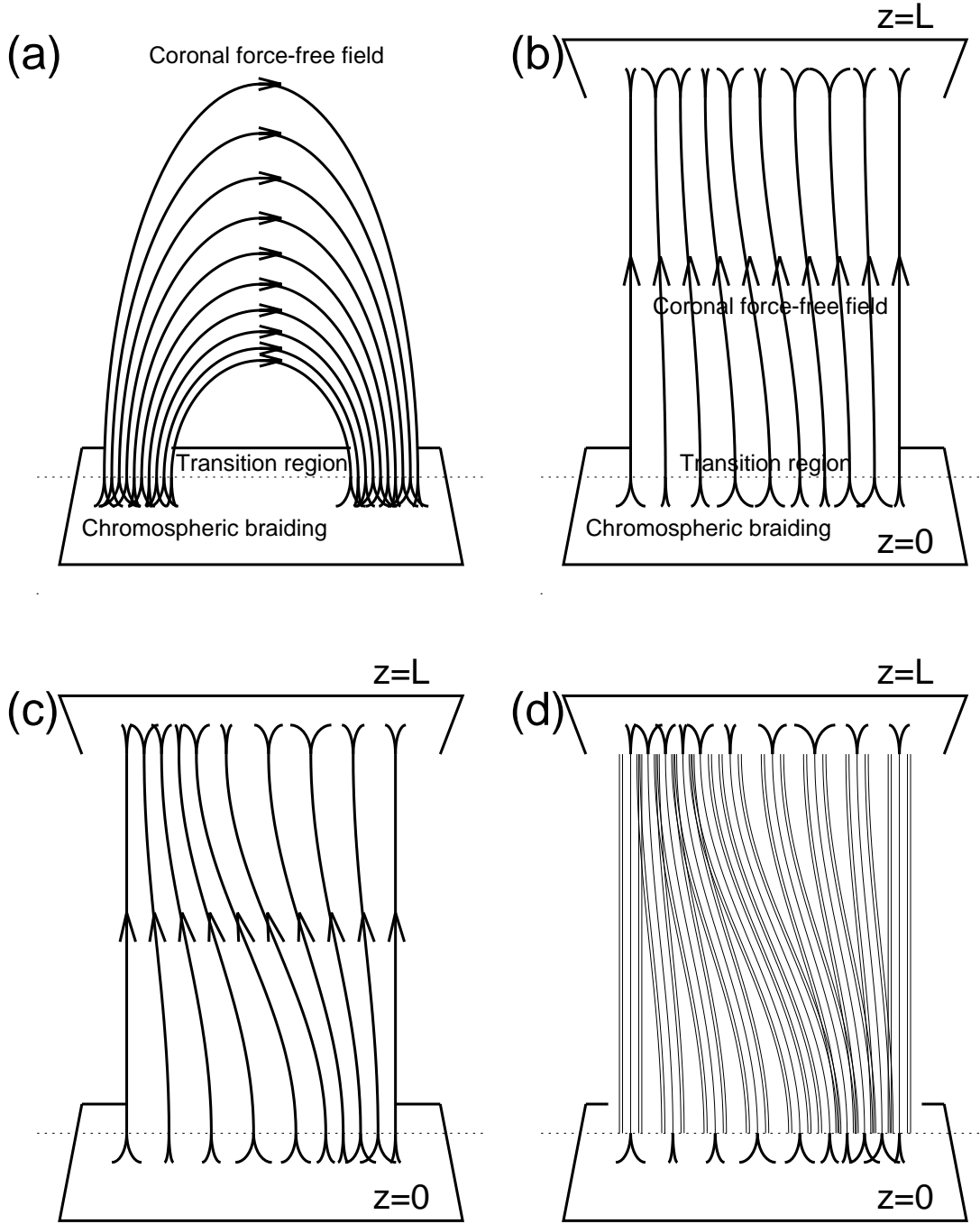}}
\caption{Revised versions of Parker's cartoons shown in Fig.~1:
Schematic drawings of field lines of a bipolar magnetic region (a),
a bipolar bundle of loop stretched out in vertical direction (b),
the same bundle with stronger twisting (c), and twisted flux tubes
after some random walk of the loop footpoints (d). Note that all
the loop braiding occurs in the chromosphere or transition region,
while the coronal parts of the loops are force-free solutions with
small misalignment angles between neighbored field lines.}
\end{figure}

\begin{figure}
\centerline{\includegraphics[width=1.0\textwidth]{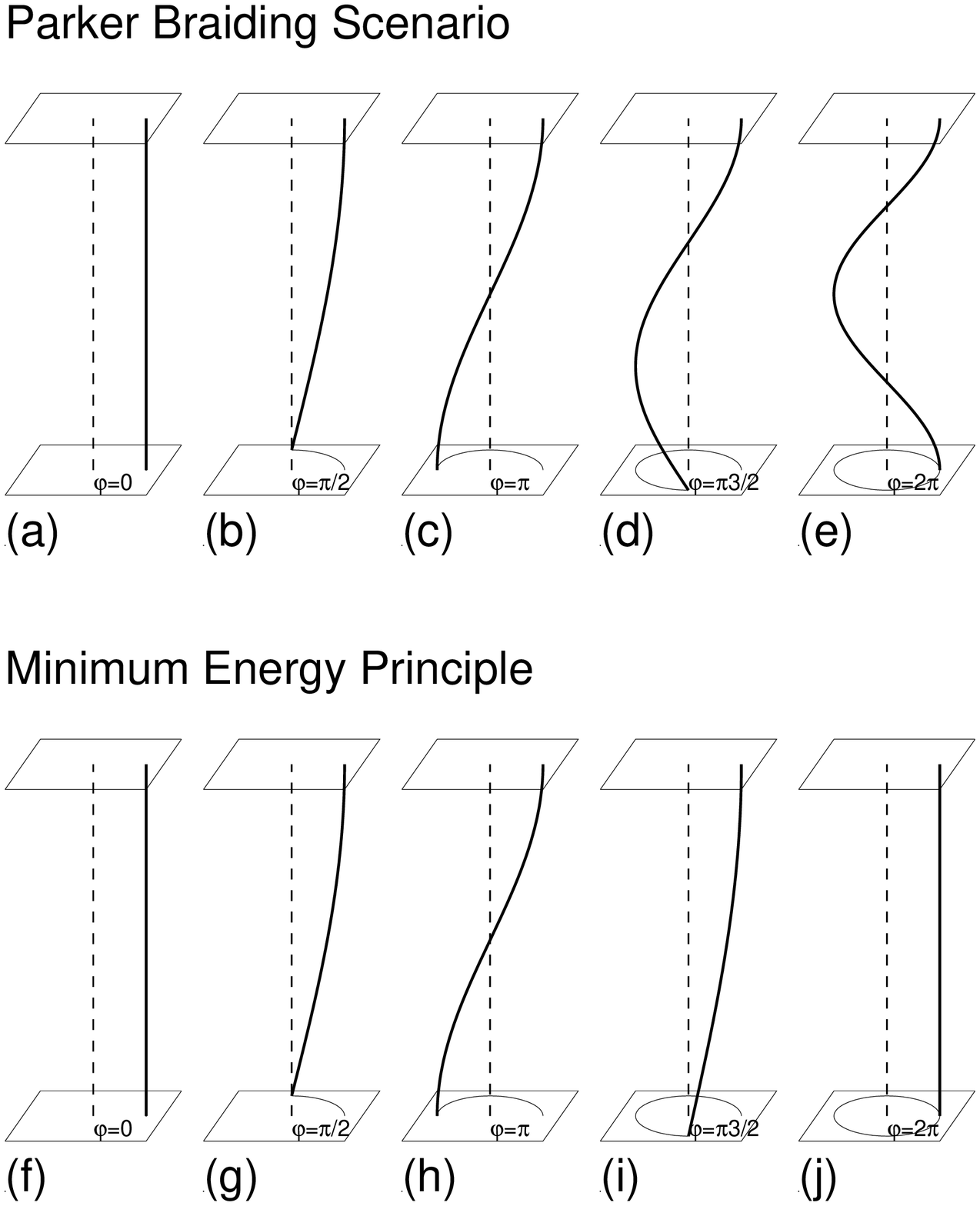}}
\caption{Helical twisting of a force-free magnetic field line
in a stretched-out geometry in Parkers' braiding scenario (a-e),
and according to the minimum energy principle (f-j). Note that
there is a bifurcation at a rotation angle of $\varphi=\pi$ (c, h),
whereafter the adjustment to a force-free solution follows
a different evolution in the two scenarios c-e and h-j.}
\end{figure}

\begin{figure}
\centerline{\includegraphics[width=1.0\textwidth]{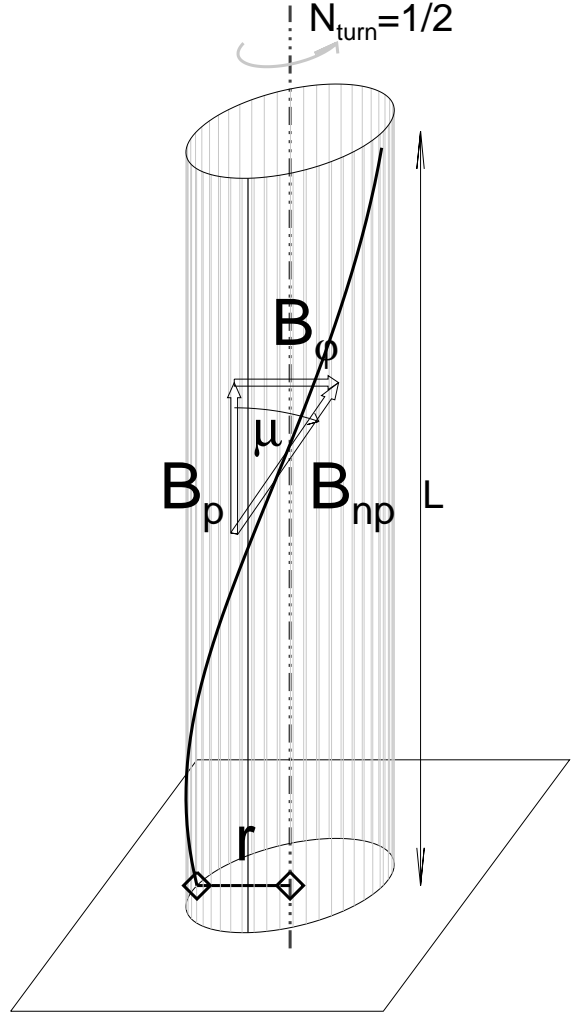}}
\caption{The basic 3-D geometry of a cylindrical flux tube with uniform
twist is defined by the length $L$ of the cylinder axis, the number
of twisting turns along this length, $N_{\rm turn}$, and by the misalignment
angle $\mu$ at the flux tube radius $r$ between the potential field line
${\bf B}_p$ (aligned with the cylindrical axis) and the non-potential
field line ${\bf B}_{np}$ (aligned with the twisted loop). The
non-potential field line ${\bf B}_{np}$ can be decomposed into a
longitudinal (potential) field component $B_p$ and an azimuthal field 
component $B_{\varphi}$.}
\end{figure}

\begin{figure}
\centerline{\includegraphics[width=1.0\textwidth]{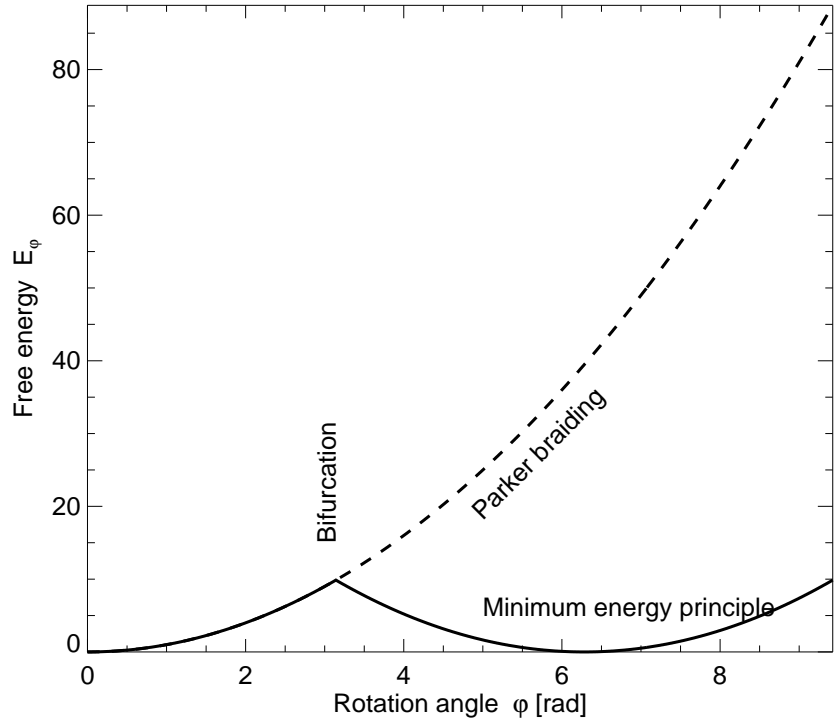}}
\caption{Dependence of the free energy of a helically twisted flux tube
for the Parker braiding scenario (dashed curve) and for the minimum
energy principle (solid line). A bifurcation occurs at $\varphi=\pi$.}
\end{figure}

\begin{figure}
\centerline{\includegraphics[width=1.0\textwidth]{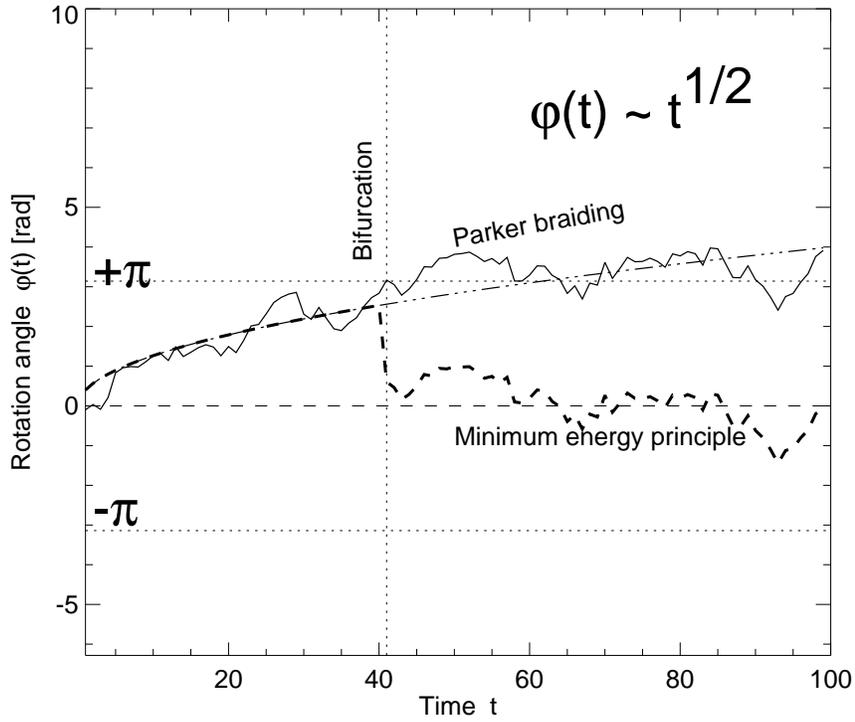}}
\caption{Diagram of the evolution of the rotation angle $\varphi(t)$
between an untwisted and twisted field line for Parker's
braiding scenario (solid curve) and for the minimum energy principle
(thick dashed curve). The diffusive random walk $\varphi(t) \propto t^{1/2}$
is indicated (dot-dashed curve) and the bifurcation point (vertical
dotted line) from where the evolution of the two models differs.
The minimum energy principle allows rotation angles with
$-\pi \le \varphi \le +\pi$.}
\end{figure}

\end{document}